# The Effective Temperature Concept Tested in an Active Colloid Mixture

Ming Han,[1,†] Jing Yan,[2,†] Steve Granick,[3*] Erik Luijten[4*]

[1]Applied Physics Graduate Program, Northwestern University, Evanston, IL 60208, USA

[2]Department of Materials Science and Engineering, University of Illinois, Urbana, IL 61801, USA

[3]IBS Center for Soft and Living Matter, UNIST, Ulsan 689-798, South Korea

[4]Departments of Materials Science and Engineering, Engineering Sciences and Applied Mathematics, and Physics and Astronomy, Northwestern University, Evanston, IL 60208, USA

[†] M.H. and J.Y contributed equally to this work.

*Thermal energy agitates all matter and its competition with ordering tendencies is one of the most fundamental organizing principles in the physical world[1]. Thus, it is natural to enquire if an effective temperature could result when external energy input enhances agitation[2,3]. Potentially this could extend the insights of statistical thermodynamics to nonequilibrium systems, but despite proposals that the effective temperature concept may apply to synthetic active matter[4,5], biological motors[6-8], granular materials[9-11] and turbulent fluids[12], its predictive value remains unclear. Here, combining computer simulations and imaging experiments, we design a two-component system of driven Janus colloids such that collisions produced by external energy sources play the role of temperature, and in this system we demonstrate quantitative agreement with hallmarks of statistical thermodynamics for binary phase behavior: the archetypal phase diagram with equilibrium critical exponents, Gaussian displacement distributions, fluctuation-dissipation relations, and capillarity. These quantitative analogies to equilibrium expectations, observed in this decidedly nonequilibrium system, constitute an existence proof from which to compare future theories of nonequilibrium, but limitations of this concept are also highlighted.*

The tug of war between randomization from temperature and ordering from enthalpy governs whether components mix or demix at equilibrium. The even larger realm of nonequilibrium systems similarly shows empirically that particles with different dynamic modes may segregate spontaneously: for example, bacteria[13], granular particles under shaking[14], and active colloids[15]. Without wishing to minimize important differences between these phenomena, we test the concept that these nonequilibrium systems might possess an effective temperature. Experimentally, we employ monodisperse colloids that are large enough to render thermal Brownian motion negligible. Designing this system to be dynamically driven, we apply external electric and magnetic fields (Fig. 1a) to Janus silica spheres with a diameter ~3 μm, one hemisphere coated with a thin Ni/Ti cap and a $SiO_2$ protective layer to render their outermost chemical makeup homogeneous. The particles settle to the bottom of the planar sample cell but remain dispersed in water. This effectively eliminates complications of solid-solid friction[16] often encountered in granular systems. The a.c. electric field applied perpendicular to this quasi-2D system causes the particles to orient their equators vertically and self-propel ("swim") with their nonmetallic hemisphere facing forwards[17], while a magnetic field, rotating at low frequency in the 2D plane, forces swimming particles to trace circular orbits with a tunable radius $R$ (Supplementary Video 1). Spontaneous symmetry breaking results. The Janus colloids dynamically differentiate into two types that orbit 180 degrees out of phase, with their nonmetallic sides facing in the opposite direction. Particles that are in-phase never collide, whereas those that orbit out of phase have opportunities to collide with each other when close by (Fig. 1b). With careful design to match the experimental conditions in their essential respects (Methods), Brownian dynamics simulations extend the experiments to otherwise-inaccessible regimes.

We explore analogies to temperature. Without rotation, oppositely swimming particles would separate into lanes to reduce collisions, a known phenomenon[18]. However, at finite $R$, lanes grow only up to a finite length (Extended Fig. 1) and continually break and reform to follow the particle orientation (Supplementary Video 2). Viscous drag dissipates the injected energy and the collisions disturbing the intrinsic circular orbits introduce fluctuations into particle motions.

To test the claim that this system mimics thermal motion, we first study the long-time behavior of particle dynamics by inspecting their positions once per rotation cycle, at symmetric composition. Hallmarks of classical Brownian motion are observed. First, the mean squared displacement is proportional to the number of cycles (Extended Fig. 2). Second, the displacement (along either $x$- or $y$-axis) follows a Gaussian distribution over three orders of magnitude of probability density (Fig. 1c), further confirmed by inspection of the kurtosis (Extended Fig. 3). We emphasize that our experiments are consistent with the simulations wherever comparison is possible, while the simulations offer better and more exhaustive statistics. Taken together, the data in Fig. 1 suggests strongly that $R$ plays the role of effective temperature.

Strong claims demand strong evidence. The textbook Langevin approach to describe thermal fluctuations in equilibrium requires random forces to satisfy several requirements[19]: (1) their (time- or ensemble-) average must vanish; (2) orthogonal components of the random forces on a particle must be independent; (3) each component of the random force must decorrelate instantaneously and have a variance proportional to temperature. This approach, mathematically equivalent to the Fokker-Planck equation[19], is widely employed in molecular dynamics simulations to control temperature. We tested these points in simulations. Condition 1 is satisfied by symmetry of the circular motion. In this overdamped system, random forces are immediately balanced by the viscous forces and manifest themselves as perturbations around the circular orbits. As shown in Extended Fig. 4, orthogonal components of the perturbations tangential and normal to the circular orbit are independent and both decorrelate at short times, within one tenth of a cycle, satisfying Conditions 2 and 3. The resulting Smoluchowski[20] equation explains the emergence of Brownian motion. Consistently, the diffusivity $D$ measured on long time scales and its value computed from the autocorrelation function of the perturbations agree quantitatively (Fig. 1d). Furthermore, the Smoluchowski equation yields a Gaussian displacement distribution with variance determined by the variance of the random forces, regardless of whether these forces are Gaussian random variables[19] (Extended Fig. 4). This explains the Gaussian displacement distribution that we observe. Moreover, $D$ is linear with the radius $R$ (Fig. 1d). Since, according to the Einstein relation, $D$ is proportional to absolute temperature, we can represent and control the effective temperature by the radius.

Further exploring analogy to thermodynamics, we note that phase separation between the two dynamical species arises because collisions between these particles of different types lead to an effective attraction between particles of the same type. Akin to the concept of osmotic pressure, there is tendency to aggregation despite the absence of an enthalpy of interaction. To quantify this, we extract the potential of mean force[21,22] $V_{PMF}$ from the radial distribution functions $g(r)$ of the two species (Fig. 2a). This procedure yields the pair potential in the intrinsic "thermal" energy unit, even though here that is not $k_BT$. As attraction strengthens with decreasing $R$, this suggests that phase segregation might ensue. Indeed, this is what we observe (Supplementary Video 3).

The analogy to phase behavior described by equilibrium thermodynamics would be most convincing if this dynamical system were to display the most subtle aspect of a phase boundary, namely a critical point with the concomitant universal singular behavior. Symmetry dictates that this point must occur in a system with equal number of A and B particles. When $R$ is decreased to a critical value $R_c = 1.84\sigma$ in the simulations (between $\sigma$ and $2\sigma$ in experiment, depending on particle density), the spatial distribution of the local order parameter $s = \phi_A - \phi_B$ ($\phi$ the composition fraction) shows self-similar patterns indicating a diverging correlation length, the hallmark of criticality (Fig. 2b and Supplementary Video 4). Consistently, we find that the susceptibility $\chi$, the variance of the local order parameter $s$, also diverges around $R_c$ in the limit of large system size (Fig. 2c). Applying finite-size scaling[23], we find that the measured susceptibility maximum $\chi_{max}$ diverges as a function of linear system size $L$ (Fig. 2c, inset) with exponent $1.71 \pm 0.05$, which agrees with the 2D Ising critical exponent ratio $\gamma/\nu = 7/4$ expected for an equilibrium system with scalar order parameter and short-range interactions. Moreover, the global order parameter (Methods, Extended Fig. 5) exhibits singular behavior $M \sim |R - R_c|^\beta$ with $\beta = 0.127 \pm 0.003$, consistent with the value $1/8$ in the 2D Ising universality class. The analogy with equilibrium critical phenomena extends beyond static properties, encompassing even dynamic effects: the relaxation time of the system diverges as $(R - R_c)^{-z\nu}$ upon approaching the critical point from the supercritical regime (Fig. 2d), with $z\nu = 3.7 \pm 0.1$, in agreement with $z\nu = 4 - \eta = 3.75$ for the 2D Ising universality with a conserved order parameter[24].

Thus, these observations quantitatively reflect all universal aspects of a continuous phase transition, including a consistent and complete set of critical exponents belonging to the equilibrium universality class. As in an equilibrium system all of these findings can be derived from a free energy functional[23,24], we conjecture that a similar framework may apply to steady-state nonequilibrium systems also, provided that the reasoning leading to designing this experimental system is satisfied. This is a decidedly nontrivial consequence of the analysis presented here.

We map a complete phase diagram parametrized by $R$ and $\phi_A$. Fig. 2f visually resembles the established phase diagram of binary fluids, sectioned by a binodal and a spinodal curve. Validating the simulations, experiments within the phase boundary display the expected interconnected patterns known as spinodal decomposition (Fig. 2e, Supplementary Video 5) at the critical composition, and nucleation-and-growth behavior (Supplementary Video 6) at an off-critical compositions close to the phase boundary. Their differences are clear at the initial stage (Fig. 3a): whereas the former displays patterns with a specific wavelength, the latter relies on random excitations to form a nucleus and hence lacks a characteristic length scale. Nascent nuclei fluctuate wildly, then merge and grow into spherical domains with time (Supplementary Video 6, 7). Late-stage coarsening behavior of the spinodal pattern is also consistent with the equilibrium concept of surface tension driving domain growth with a power law exponent of 0.5 (Fig. 3b-d and Extended Fig. 6). The existence of an effective surface tension can be clearly seen from the spreading of a spherical droplet onto a larger domain (Fig. 3b and Supplementary Video 8). However, the capillary velocity (Fig. 3d, inset) is seven orders of magnitude less than that of water, underscoring a significant quantitative difference from equilibrium behavior despite the display of similar generic patterns. Spinodal patterns at the critical composition, and nucleation patterns off the critical composition, present another pleasing analogy to equilibrium mixtures.

We conclude with cautionary comments about limitations of the effective temperature concept. First, the mixed state at high effective temperature, the so-called "one-phase" region, consists of lanes (Supplementary Video 2). Second, collisions at the domain boundary during coarsening drive particles to move in opposite directions parallel to the boundary, unlike an equilibrium system (Extended Fig. 7). Third, once phase separation sets in, the effective temperature drops to zero locally and these systems crystallize (Supplementary Videos 6-8), unlike equilibrium systems where temperature is inherently uniform. The findings in this paper should generalize naturally to other dynamic components that perform 3D periodic or chaotic motion that traces back to the origin with statistically significant numbers of collisions. Yet these limitations underscore that care is needed to apply the effective temperature concept despite the nontrivial predictive capability demonstrated here.

## Methods

**Particle Synthesis:** Onto a planar submonolayer of monodisperse 3 μm silica particles (Tokuyama), obtained by drying 20 μL 2 wt% particle suspension in deionized (DI) water onto a half glass slide (1.5 inch × 1 inch), a thin film of 4 nm Ni/10 nm Ti/5 nm $SiO_2$ is sequentially deposited vertically using electron-beam deposition in vacuum. The monolayer is washed thoroughly with DI water and isopropyl alcohol, and then sonicated in 20 mL DI water to collect the particles in a 50 mL centrifuge tube.

**Setup:** A spatially homogeneous rotating magnetic field is generated by two orthogonal pairs of solenoids as described previously[25]. Briefly, the two pairs of solenoids receive two sinusoidal voltages from a function generator, amplified by power amplifiers, with $\pi/2$ phase difference between each signal to produce an in-plane rotating field $B$.

The electric field is applied by sandwiching the particles between two coverslips coated with indium tin oxide (ITO) from SPI[26]. The ITO-coated coverslips are further coated with 25 nm $SiO_2$ using electron-beam deposition to electrostatically prevent the particles from surface adsorption. A thin strip on one side of the ITO-coated coverslip is left uncoated during the $SiO_2$ deposition, which is used later to connect to a function generator (Agilent 33522A) via copper tapes. A square wave of 5 kHz and 7 V is used for all experiments. The two ITO-coated coverslips are separated by a spacer (Grace Biolab, SecureSeal) about 120 μm thick, with a 9 mm diameter hole.

Movies are taken in a customized microscope using an LED light source (Thorlabs MCWHL2) and a CMOS camera (Edmund Optics 5012M GigE). For a high-magnification view in which the Janus features can be resolved, a 50× long-working-distance objective (Mitutoyo, numerical aperture 0.55) is used. For large-scale views, a 10× long-working-distance objective (Mitutoyo, numerical aperture 0.28) is used. In this case, the light source is intentionally misaligned with the optical axis, such that the metal coating casts a shadow on the particle. Hence, particles that instantaneously face their metal side towards the light source appear darker than those facing the opposite direction. We then use stroboscopic sampling, only analyzing one frame per cycle, such that one species is consistently brighter than the other.

**Experimental procedure:** 10 µL of concentrated particle suspension is dropped onto an ITO-coated coverslip with spacer, and then sandwiched with another ITO coverslip. This produces a monolayer of colloids with area fraction between 0.35 and 0.45. Repeated experiments can be done on one sample for 2 to 3 hours, after which significant particle clustering or substrate adsorption occurs. Before each run, a constant DC magnetic field is applied in the image plane to cause particles to assemble into zig-zag chains, which ensures a completely mixed initial state without memory of earlier states. A rotating magnetic field is then applied with strength of 5 mT and frequency ranging from 16 Hz to 0.05 Hz, followed immediately by an a.c. electric field (5 kHz, 58 V/mm). This we define as $t = 0$ s. Movies are taken up to 20 minutes for each condition, at a rate of 16 frames per second.

To change the ratio between the two species, a pulse of additional $z$-axis magnetic field is applied. Although the precise mechanism of the conversion is unclear, we are able to roughly control the composition by varying the strength and duration of this $z$-field.

**Image processing for large view:** We first subtract background light gradient due to the optical misalignment, obtained by averaging over a whole movie. The image is then smoothened and binarized, with each pixel now having a local order parameter $s(r)$ (defined below) of value +1 or

−1. Spatial correlation $C(r)$, defined as $\langle s(0) \cdot s(r) \rangle$, is calculated from the binarized image. Correlation length $\xi$ is defined as the $r$ value where $C(r)$ first passes zero. Linear interpolation between the last positive value and first negative value is used to improve accuracy. Structure factor is obtained by applying fast Fourier transform to the image and taking a radial average.

**Simulation procedure:** We model this overdamped and fully driven quasi-2D system via athermal Brownian dynamics simulations. Confined to the $x$–$y$ plane, Janus particles of diameter $\sigma = 3$ μm with area fraction 0.37 (to match the experiment) move in a square, periodic domain of side length $200\sigma$. Their positions $\boldsymbol{r}(t)$ and orientations $\boldsymbol{\Omega}(t)$ follow the master equations

$$\left\{ \begin{array}{l} \dot{\boldsymbol{r}}(t) = \boldsymbol{F}/\zeta_{\mathrm{t}} \\ \dot{\boldsymbol{\Omega}}(t) = \boldsymbol{T} \times \boldsymbol{\Omega}/\zeta_{\mathrm{r}} \end{array} \right. ,$$

where $\boldsymbol{F}$ and $\boldsymbol{T}$ are the force and the torque exerted on the particle, respectively, and $\zeta_{\mathrm{t}}$ and $\zeta_{\mathrm{r}}$ refer to the translational and rotational drag coefficients. Here reduced units are employed, with the particle mass $m = 3.74 \times 10^{-14}$ kg as mass unit, the particle diameter $\sigma$ as length unit, and $\tau = 10^{-3}$ s as time unit. Given the dynamic viscosity $\mu$ of the water, those spherical particles have $\zeta_{\mathrm{t}} = 3\pi\mu\sigma = 672.8$ and $\zeta_{\mathrm{r}} = \pi\mu\sigma^3 = 224.3$. According to the reciprocal theory[27], the action of the electric and magnetic fields in experiment can be modeled as a driven circular motion caused by an external force $F_{\mathrm{ex}} = \zeta_{\mathrm{t}}v = 4.7$ (with $v \approx 21$ μm/s the measured terminal swimming velocity) and an external torque $T_{\mathrm{ex}} = \zeta_{\mathrm{r}}\omega$ (with $\omega$ the tunable angular frequency). Two types of particles are considered, with oppositely directed driving forces at each instant.

In addition to the external driving, interactions between the particles also affect their motions. Excluded-volume effects are implemented as shifted-truncated Lennard-Jones interactions (STLJ) with $\sigma_{LJ} = 1$, $\varepsilon_{LJ} = 1$, and cutoff at $r_c = 2^{1/6}$. In experiment, due to the $z$-directional electric field, the particles experience an induced dipolar force, which is isotropic in the $x$–$y$ plane and decays as $A \cdot r^{-4}$. Here we set the strength $A = 7.5$ so that the dipolar interaction balances the driving force at $r_c$.

All simulations start with a random configuration and proceed for at least $10^9$ steps with time step $dt = \tau$. This corresponds to around $10^5$ rotation cycles (equivalent to 50 hours in experiment), which we found to be sufficient for the system to reach a steady state.

**Analysis of phase transition:** To quantify the phase transition, we define the order parameter as the composition difference $s = \phi_A - \phi_B$, which is conserved globally but fluctuates locally in the system. Here we divide the entire system into relatively small subboxes of side length $L = 12.5\sigma$ and record their composition fluctuations, which mimic those present in the grand-canonical ensemble. The histogram yields the order-parameter distribution, which changes from a Gaussian distribution centered at 0 to a symmetric bimodal distribution as $R$ decreases at symmetric composition (Extended Fig. 5a). The peak position denotes the global order parameter $M$ of the system, specifically,

$$M = \begin{cases} s_{\text{peak}}, & \text{single peak } s_{\text{peak}} \\ \left(\left|s_{\text{peak}}^-\right| + \left|s_{\text{peak}}^+\right|\right)\big/2, & \text{two peaks } s_{\text{peak}}^- \text{ and } s_{\text{peak}}^+ \end{cases}.$$

It displays singular behavior at the transition point $R_c$ (Extended Fig. 5b). The susceptibility is defined as

$$\chi = \begin{cases} L^2 \left\langle \left(s - s_{\text{peak}}\right)^2 \right\rangle, & \text{single peak } s_{\text{peak}} \\ L^2 \left[ \left\langle \left(s - s_{\text{peak}}^-\right)^2 \right\rangle_{s \leq s_{\text{peak}}^-} + \left\langle \left(s - s_{\text{peak}}^+\right)^2 \right\rangle_{s > s_{\text{peak}}^+} \right], & \text{two peaks } s_{\text{peak}}^- \text{ and } s_{\text{peak}}^+ \end{cases}.$$

To study the divergence of $\chi$, we perform a finite-size scaling analysis by systematically varying the subbox size $L$ from $5.6\sigma$ to $20\sigma$ and exploring the rise of the maximum susceptibility $\chi_{\text{max}}$ with $L$ (Fig. 2c).

## Acknowledgements

At the IBS Centre for Soft and Living Matter, SG acknowledges support by the Institute for Basic Science, project code IBS-R020-D1. This work was supported by the U.S. Department of Energy, Division of Materials Science, under Award DE-FG02-07ER46471 through the Frederick Seitz Materials Research Laboratory at the University of Illinois at Urbana-Champaign (J.Y. and S.G.) and by the National Science Foundation under Award Nos. DMR-1121262 and DMR-1610796 (M.H. and E.L.). We acknowledge support from the National Science Foundation, CBET-0853737 for equipment and from the Quest high-performance computing facility at Northwestern University (M.H. and E.L.). J.Y. holds a Career Award at the Scientific Interface from the Burroughs Wellcome Fund.

## Author Contributions

J.Y. and S.G. initiated this work. J.Y. performed the experiments; M.H. and E.L. performed the modeling and simulations. J.Y., M.H., E.L. and S.G. wrote the paper.

## Competing interests statement

The authors declare no competing financial interests.

# Figures

Figure 1

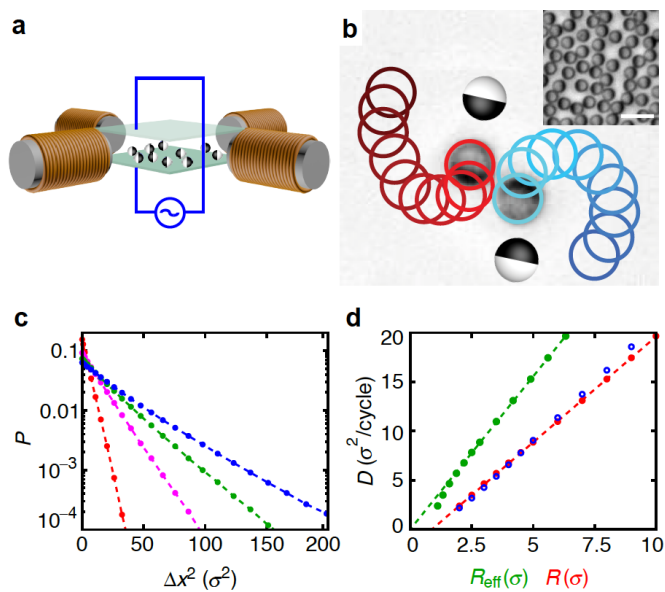

**Figure 1. Effective temperature. a,** A sketch of the experimental setup. A vertical a.c. electric field and a planar rotating magnetic field are applied to a suspension of Janus colloids. **b,** Experimental image of a two-particle collision event overlaid with reconstructed trajectories, colored-coded with time (varying from dark to bright). Inset: Experimental image of oppositely moving particles. These particles align their Janus interfaces (separating the metal and the silica hemispheres) with the present magnetic field to minimize energy, while spontaneously separating into two species with their nonmetallic hemispheres facing in opposite directions. All particles perform circular motion with radius $R = 3.52\sigma$ ($\sigma = 3$ μm the particle diameter). Scale bar: 10 μm. **c,** Probability of displacement $\Delta x$ along the $x$-axis after one rotation cycle, measured in simulations with $R = 2\sigma$ (red), $4\sigma$ (purple), $6\sigma$ (green) and $8\sigma$ (blue). **d,** Dependence of diffusivity $D$ on $R$. Direct measurement and prediction based on instantaneous particle collisions are shown as filled and open circles, respectively. The dependence is evaluated with respect to both the intrinsic radius $R$ (red) as determined only by the rotating magnetic field and the effective radius $R_{\text{eff}}$ (green) modulated by particle collisions.

Figure 2

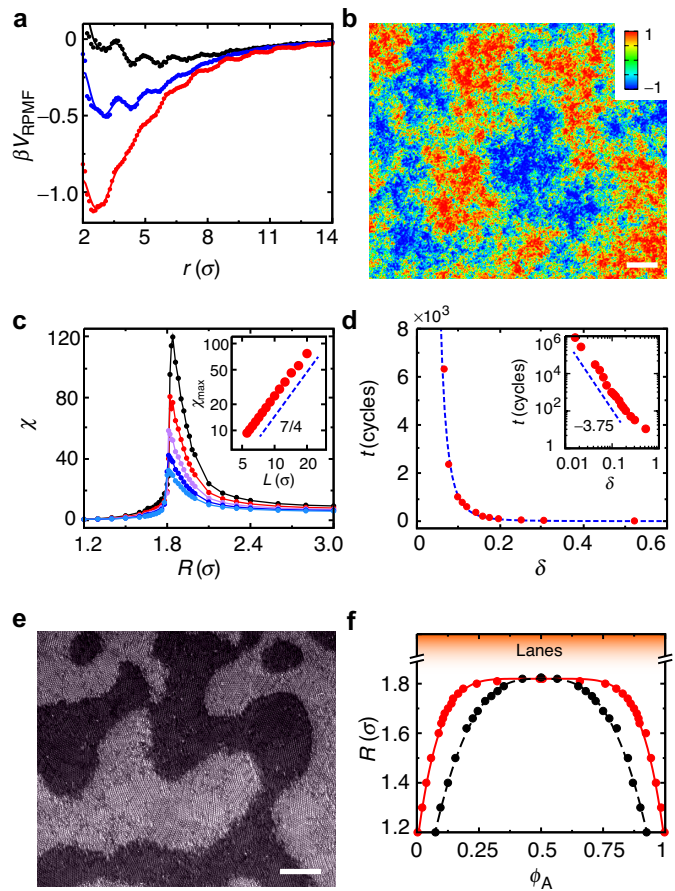

**Figure 2. Critical phenomena and phase separation. a,** Relative potential of mean force $V_{RPMF}$ at $R = 1.5\sigma$ (red), $3\sigma$ (blue) and $4.5\sigma$ (black). **b,** Spatial distribution of the local order parameter $s$ in simulation at the critical point $R_c = 1.84\sigma$. Scale bar: $120\sigma$. **c,** Dependence of susceptibility $\chi$ measuring the composition fluctuations on $R$ for a symmetric mixture, evaluated for subcells of different side lengths $L = 16.67\sigma$ (black), $12.5\sigma$ (red), $10\sigma$ (purple), $8.33\sigma$ (navy) and $7.14\sigma$ (blue). Inset: Relation between maximum susceptibility $\chi_{max}$ and subcell size $L$, plotted on a logarithmic scale. The blue line represents the 2D Ising power-law 7/4. **d,** Correlation time of the entire system as a function of $\delta = (R - R_c)/R_c$. Inset employs a log-log scale. The blue curve has the universal power-law exponent $-15/4$. **e,** Experimental image at $R = 0.88\sigma$. Scale bar: 60 μm.

**f,** Simulated phase diagram. Binodal and spinodal curves are shown in red and black, respectively.

Figure 3

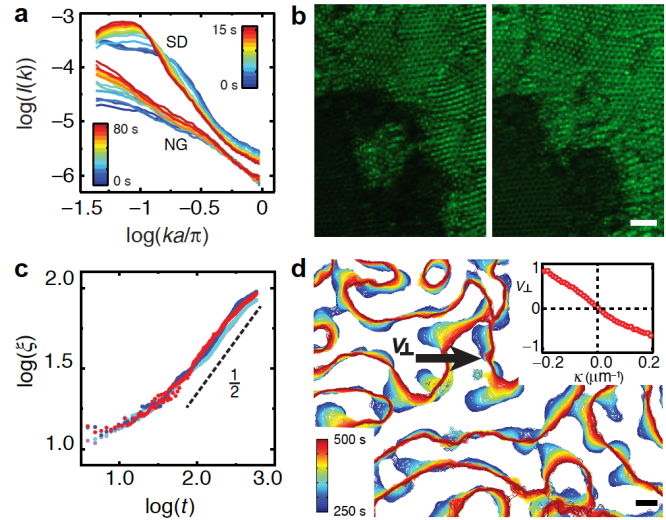

**Figure 3. Dynamics of phase segregation and coarsening. a,** Time evolution of the scattering function $I(k)$ at initial states, for a system undergoing spinodal decomposition (SD; $\phi_A = 0.5$, $R = 0.88\sigma$) as well as a system exhibiting nucleation and growth (NG; $\phi_A = 0.15$, $R = 0.5\sigma$). **b,** Sequential experimental images showing a spreading process. Time difference: 42 s. Scale bar: 20 μm. **c,** Time evolution of the correlation length $\xi$. Four independent samples are shown. **d,** Time evolution of domain boundaries. Inset: Normal component of boundary velocity $V_\perp$ as marked in the main figure is plotted with respect to local curvature $\kappa$. Scale bar: 50 μm.

## Extended Figures

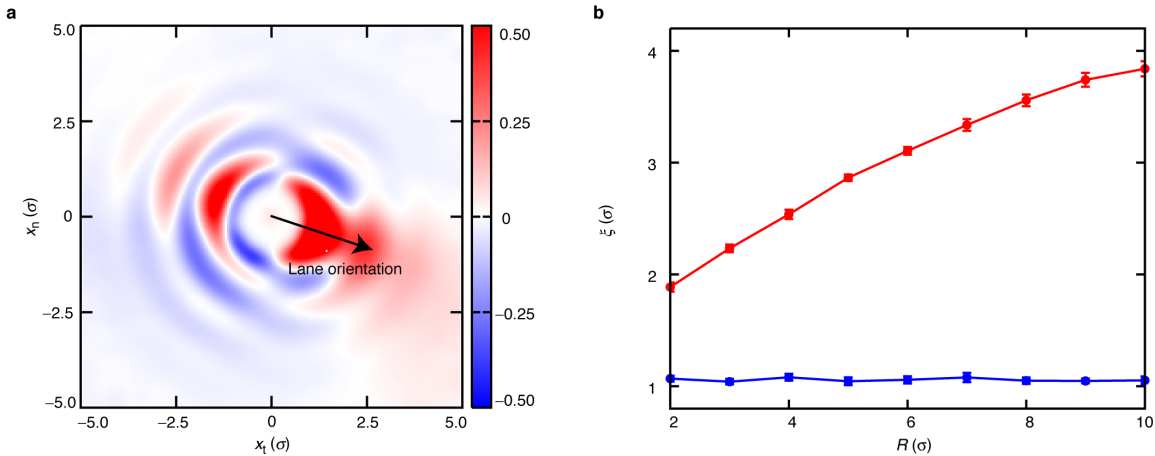

**Extended Figure 1. 2D spatial correlation characterizing lane formation. a.** Spatial correlation $C(x_t, x_n)$ at radius of rotation $R = 8\sigma$, defined as $\langle s(0, 0) \cdot s(x_t, x_n) \rangle$ in the particle reference frame, where $(x_t, x_n)$ stands for coordinates tangential and normal to the particle orientation. Particles exhibit strong correlations only along the particle orientation, indicating lane formation. In fact, the formed lanes lie themselves in a small angle lagging behind the particle orientation, as the reorientation of the lanes generally requires multiple particle collisions and reorganization. **b.** Dependence of corresponding correlation length $\xi$ on $R$. Correlation lengths are evaluated along (red) and perpendicular (blue) to the lane orientation, respectively. They are denoted as the distance between the origin and the position where the correlation $C$ becomes less than 0.1. As $R$ increases, lanes become longer but remain narrow. Note that $\xi$ is the correlation length of the local order parameter, rather than that of the order-parameter fluctuations discussed in the main text. The latter diverges at the critical point $R_c$.

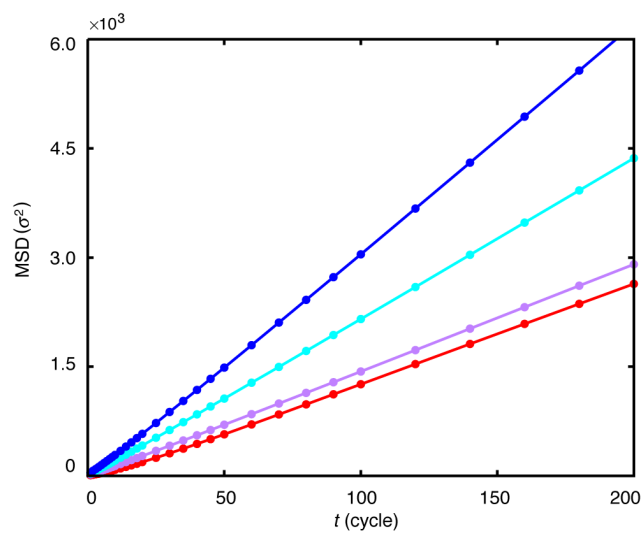

**Extended Figure 2. Mean squared displacement at different *R*.** Mean squared displacement (MSD) displays a linear dependence on stroboscopic time (in unit of cycles), at all investigated radius of rotation $R = 1.84\sigma$ (red), $3\sigma$ (purple), $4.5\sigma$ (cyan) and $6\sigma$ (blue). This linearity indicates that the long-time dynamics of the particles are diffusive.

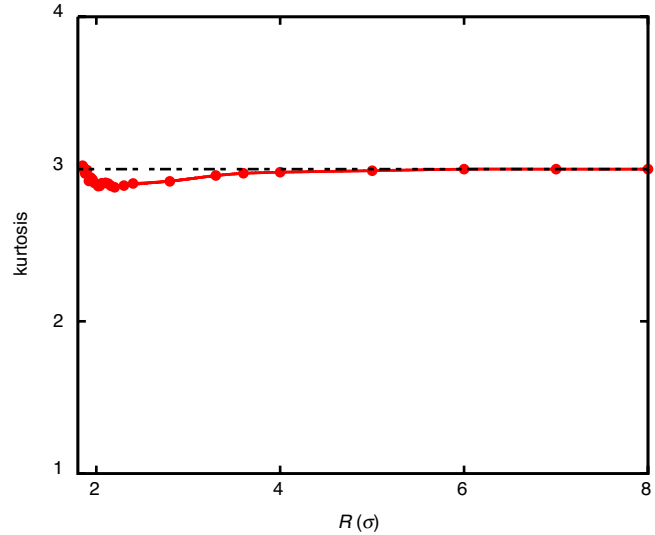

**Extended Figure 3. Kurtosis versus *R*.** Kurtosis, $\langle \Delta x^4 \rangle / \langle \Delta x^2 \rangle^2$, of the distribution of stroboscopic displacement $\Delta x$ along the *x*-axis (shown in Fig. 1c) at various radii of rotation *R*. The kurtosis, which has a lower bound of 1, is close to its Gaussian value 3 for all *R* investigated, indicating that particle displacement follows a normal distribution. Minor suppression of the kurtosis below 3 occurs when *R* is decreased below $4\sigma$, since then the number of collisions per orbit becomes too small to generate large displacements. Ultimately, the kurtosis increases again due to competing effects driven by the presence of a critical point just below $R = 2\sigma$.

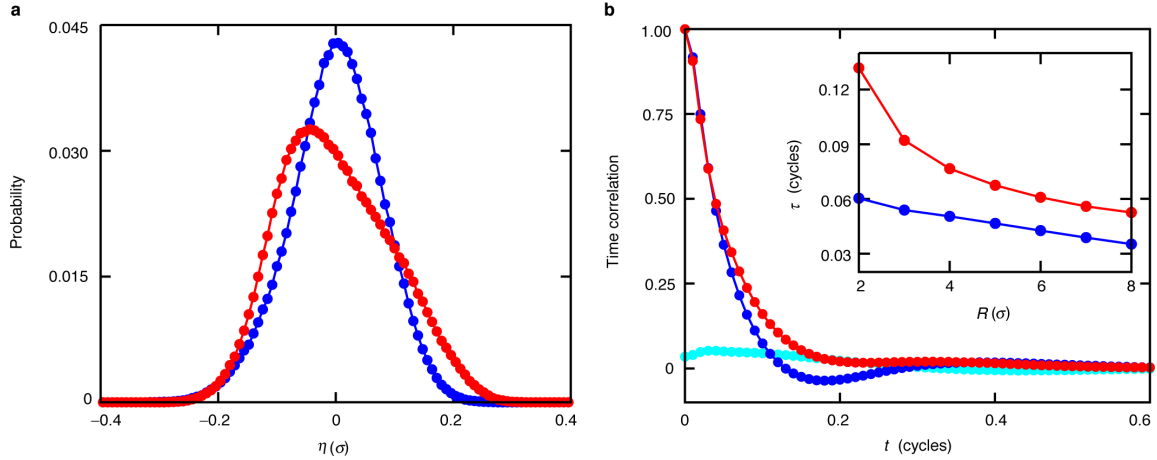

**Extended Figure 4. Statistics of collision-induced perturbations.** In this overdamped system, random forces are immediately balanced by friction with the solvent but manifest themselves as instantaneous perturbations to the circular motion. Such short-time dynamics are studied by recording particle displacements $\Delta\boldsymbol{x}$ with time interval $\Delta t = 0.01$ cycles in the particle reference frame. The displacements are further decomposed into the average $\langle\Delta\boldsymbol{x}\rangle$ due to the circular motion and the remaining fluctuations $\boldsymbol{\eta}$ corresponding to random perturbations. **a.** Probability distributions of perturbations tangential ($\eta_t$, in red) and normal ($\eta_n$, in blue) to the circular motion at $R = 4\sigma$. Both $\eta_t$ and $\eta_n$ are non-Gaussian random variables but with vanishing averages, namely $\langle\eta_t\rangle = \langle\eta_n\rangle = 0$. **b.** Autocorrelation functions of $\eta_t$ (red) and $\eta_n$ (blue) as well as their cross correlation function (cyan). Both $\eta_t$ and $\eta_n$ quickly decorrelate with themselves, and they are also almost independent with each other at all times. Inset: Dependence of relaxation time $\tau$ on $R$ for both $\eta_t$ (red) and $\eta_n$ (blue). For most of the investigated $R$, perturbations decorrelate within one tenth of a cycle, which is much smaller than the stroboscopic time scale (in units of cycles). Therefore, the particles perform 2D Brownian motion in their own reference frames. This leads to the particle diffusion in the lab reference frame as well. The corresponding diffusivity $D$ can be indeed predicted by the autocorrelation functions of the perturbations $\eta_t$ and $\eta_n$. Specifically, $D = \sum_{i=0}^{\infty} \Big[ \eta_t(0) \cdot \eta_t(i\Delta t) + \eta_n(0) \cdot \eta_n(i\Delta t) \Big] \cdot \cos(i\omega\Delta t) \cdot \Delta t$, where $\cos(i\omega\Delta t)$ is the curvature correction caused by the circular motion. As shown in Fig. 1d, this prediction matches the measured $D$.

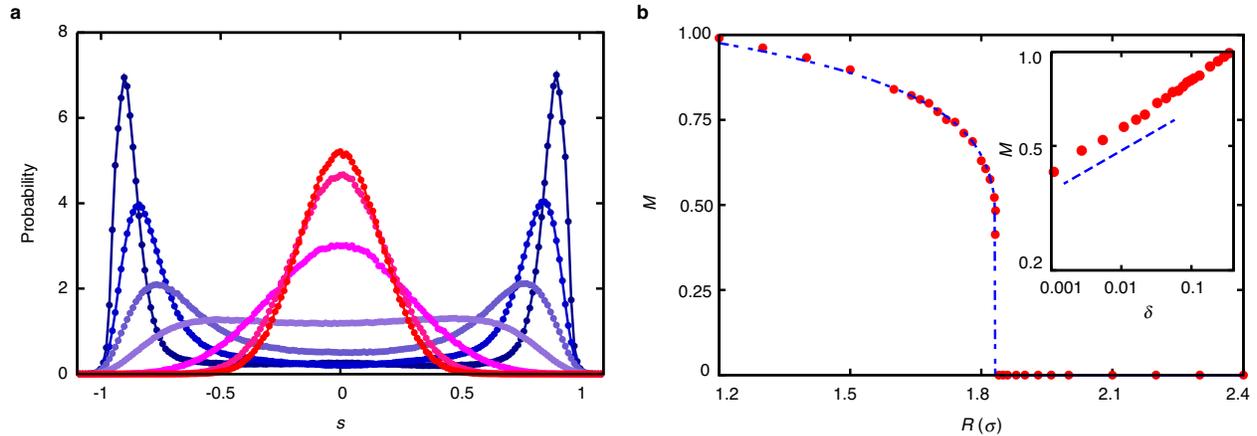

**Extended Figure 5. Order parameter characterizing phase transition. a.** Distribution of the local order parameter $s = \phi_A - \phi_B$ measured in subboxes of side length $L = 12.5\sigma$, at different $R = 1.2\sigma$, $1.4\sigma$, $1.6\sigma$, $1.8\sigma$, $2.0\sigma$, $2.2\sigma$, $2.4\sigma$ and $3.0\sigma$ (color: from blue to red), at symmetric composition. As $R$ decreases, the order-parameter distribution changes from a Gaussian distribution centered at 0 to a symmetric bimodal distribution, indicating a phase transition and the associated spontaneous symmetry breaking in the order parameter. The absolute value of the peak position denotes the global order parameter $M$ of the system. **b.** Dependence of the global order parameter $M$ on $R$. It exhibits singular behavior around the transition point $R_c = 1.84\sigma$. Inset: $M$ versus $\delta = (R_c - R)/R_c$ (for $R < R_c$) on a logarithmic scale. The blue line corresponds to a power-law exponent of $\beta = 1/8$, reflecting the order parameter critical exponent in the 2D Ising universality class.

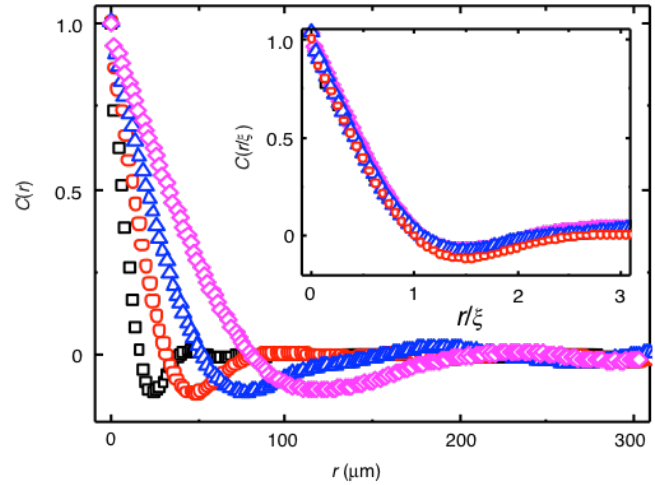

**Extended Figure 6. Dynamics of spatial correlation function $C(r)$ characterizing domain coarsening in experiment.** Spatial correlation function $C(r)$, defined as $\langle s(0) \cdot s(r) \rangle$, calculated at different times $t = 25$ (black), 100 (red), 250 (blue) and 500 s (purple), corresponding to data presented in Fig. 3c. The correlation length $\xi$ can be approximated as the distance $r$ where $C(r)$ first passes zero, and its time evolution is shown in Fig. 3c. Inset: by normalizing the distance $r$ by the time-dependent correlation length $\xi(t)$, spatial correlation functions at different time collapse, indicating that the system evolves in a self-similar manner over time.

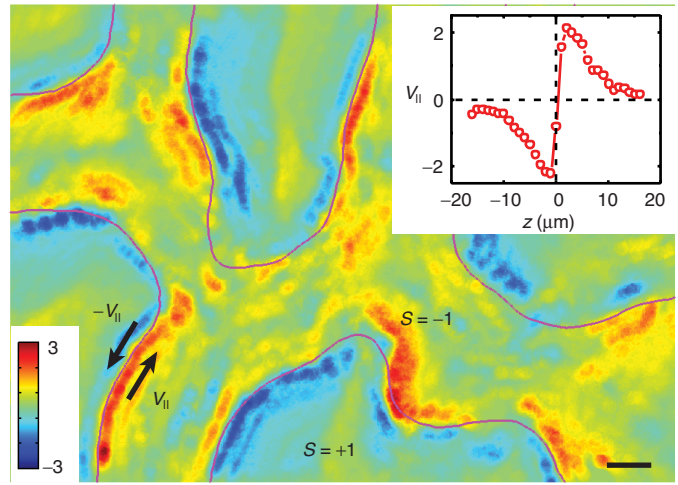

**Extended Figure 7. Particle velocity distribution at $R = 0.54\sigma$ in experiment.** A heat map of velocity $V_{\parallel}$, defined as particle velocity parallel to the nearby domain boundary and averaged over 5 s during which the domain boundary barely changes, is shown. Unit: μm/s. Scale bar: 10 μm. Due to collisions, particles on the two sides of a domain boundary move in the opposite direction. Inset: dependence of $V_{\parallel}$ on the distance $z$ to the domain boundary. Away from the boundary, $V_{\parallel}$ first displays a linear decay, akin to a standard viscous response. Therefore, one can conceptually think of the two colliding domains as shearing each other. Towards the interior of the domain, the velocity levels off, consistent with the observed crystallization.